\documentclass[aps,prd,reprint,superscriptaddress,longbibliography,nofootinbib]{revtex4-2}
\usepackage[T1]{fontenc}
\usepackage{lmodern}
\usepackage[utf8]{inputenc}
\usepackage[english]{babel}
\usepackage{amsmath,amssymb,amsthm,bm,mathrsfs}
\usepackage{graphicx}
\usepackage[colorlinks=true,linkcolor=blue,citecolor=blue,urlcolor=blue]{hyperref}
\newcommand{\dd}{\mathrm d}
\newcommand{\ii}{\mathrm i}
\newcommand{\ee}{\mathrm e}

\newtheorem{theorem}{Theorem}

\theoremstyle{definition}

\begin{document}
\title{Modular fluctuation relations on local causal horizons}
\author{Marcos L. W. Basso\,\href{https://orcid.org/0000-0001-5456-7772}{\includegraphics[height=8pt]{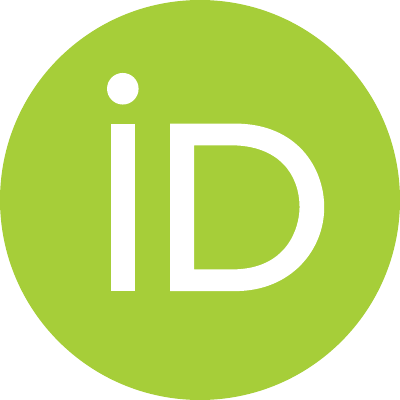}}}
\email{marcoslwbasso@hotmail.com}
\affiliation{Department of Applied Mathematics, State University of Campinas, 13083-859 Campinas, S\~ao Paulo, Brazil}
\affiliation{Nordita, KTH Royal Institute of Technology,
Stockholm University, and Uppsala University,
SE-106 91 Stockholm, Sweden}
\begin{abstract}
Local causal horizons provide a setting for studying
horizon thermodynamics in the neighborhood of any
regular spacetime point. Here, we investigate this
description beyond mean energy and entropy balances
by studying fluctuations of boost-energy transfer
induced by unitary perturbations of the horizon
reference state. Using the modular structure of the horizon algebra and a thermal reference state, we
construct an energy transfer probability measure and derive integral and detailed fluctuation relations with the mean entropy production being expressed in terms of the Araki relative entropy.
The formulation accommodates type-III algebras
without introducing a reduced density matrix and can be extended to global bifurcate Killing horizons admitting
the required thermal reference state.
For coherent scalar-field perturbations, we obtain
a closed form for the characteristic function, identify the mean transfer with the excess boost-energy flux, and derive a thermodynamic uncertainty relation bounding the
variance in terms of the mean transfer.
We then explore the geometric implications by
postulating a stochastic horizon entropy--area
correspondence that associates an area change with
each counted energy transfer. Under this additional hypothesis, we obtain fluctuation relations for area changes and an area change uncertainty bound for coherent
perturbations.
\end{abstract}
\maketitle

\section{Introduction}\label{sec:intro}

The study of horizons revealed a close connection between general relativity and thermodynamics. Hawking showed that the horizon area of a black hole cannot decrease in classical processes~\cite{Hawking71}, a result analogous to the non decrease of entropy. This correspondence was subsequently extended through the formulation of the laws of black hole mechanics~\cite{Bardeen} and the relationship was soon deepened by Bekenstein~\cite{Bekenstein1973}, who approached black hole physics from the perspective of information theory and proposed that an entropy proportional to its area should be attributed to the black hole horizon. Hawking's subsequent discovery of black-hole radiation~\cite{Hawking1975} established their thermal character and fixed the entropy--area relation. 

A related thermal phenomenon occurs even in flat spacetime. Building on Fulling's~\cite{Fulling1973} analysis of field quantization in accelerated frames, Unruh~\cite{Unruh1976} showed that a uniformly accelerated detector in the Minkowski vacuum exhibits a thermal response. The Bisognano--Wichmann theorem~\cite{BisognanoWichmann1975} provides an algebraic formulation of this thermality: under its hypotheses, the modular automorphism group of
the Rindler-wedge von Neumann algebra in the Minkowski
vacuum representation coincides, up to a fixed
rescaling of its parameter, with the Lorentz-boost
evolution preserving the wedge. Consequently, the vacuum restricted to the wedge satisfies the Kubo--Martin--Schwinger (KMS) equilibrium condition with respect to boost evolution.

The thermodynamic character of Rindler (causal) horizons suggests that this connection may also constrain spacetime dynamics locally. Jacobson~\cite{Jacobson1995} developed this idea by deriving the Einstein equations from the Clausius relation, $\delta Q=T_{\mathfrak{h}}\delta S$, imposed on all local causal horizons through each spacetime point. In this construction, $\delta Q$ is the matter boost-energy flux across the horizon, $T_\mathfrak{h}$ is the Unruh temperature, and $\delta S$ is determined by the Bekenstein--Hawking area law. This thermodynamic perspective on causal horizons \cite{JacobsonParentani2003, Padmanabhan2005, Padmanabhan2010} has motivated studies of nonequilibrium entropy production, gravitational dissipation, mechanical laws, and entanglement equilibrium~\cite{Eling2006, ChircoLiberati2010, Guedens2012a, BianchiSatz2013, Jacobson2016}. The local nature of these ideas allows their application to be explored even in spacetimes without a global bifurcate Killing horizon, motivating constructions based on others null surfaces and stretched
light cones~\cite{BaccettiVisser2014, DeLorenzoPerez2018, Parikh}. More recently, Dorau and Much related coherent
quantum-field perturbations on a local causal horizon to relative entropy and, through an entropy--area assumption, to semiclassical
gravitational dynamics \cite{DorauMuch2026}.

The thermodynamic interpretation of horizons has also
motivated the study of fluctuations beyond mean energy
and entropy balances. Iso, Okazawa, and Zhang
\cite{Iso2011} obtained Crooks- and Jarzynski-type
fluctuation relations~\cite{Jarzynski1997, Crooks1998, Crooks1999, Jarzynski11, Esposito2009, Seifert2012, Strasberg2022} that imply the generalized second law on average for black hole spacetimes~\cite{Bekenstein74, Wall2012}. These results
are complemented by the effective stochastic
description of near-horizon fields developed in
Ref.~\cite{IsoOkazawa2011}. Other approaches study entropy production and fluctuation relations for gravitational algebras \cite{CiraficiNESS2024,CiraficiFT2024}, or work statistics in AdS/CFT \cite{Takeda2026}.  Relativistic fluctuation theorems have also been developed for localized quantum systems in curved spacetime~\cite{Basso2025} and within general-relativistic stochastic dynamics~\cite{Cai25}. For quantum fields, work distributions and the first law have been formulated through local operations~\cite{Ortega,Bonfill}. Related developments address static curved spacetimes \cite{Costa2026} and field-energy exchange with an Unruh-DeWitt detector~\cite{BassoSaa2026}.

Algebraic quantum statistical mechanics provides a
framework for energy counting statistics without
requiring a trace-class density matrix. Relative
modular operators encode counting statistics
for infinitely extended systems
\cite{JOPP2011,JOPS2012}, while relative entropy
remains well defined for type-III local
algebras~\cite{Araki1976, Witten18}. Here, we connect modular counting statistics to
boost-energy transfer across local causal horizons
by using the geometric interpretation of modular
evolution as a boost flow. From this identification,
we derive fluctuation relations for local causal horizons.
The resulting construction broadens the geometric setting of energy fluctuations beyond black-hole spacetimes. 
Specifically, we consider the horizon von Neumann algebra of a local Rindler reference theory, equipped with a faithful normal state that satisfies the KMS condition with respect to the boost flow at the Unruh temperature. A unitary operation within this algebra perturbs the reference state while the
reference boost evolution is kept fixed. These
ingredients define the boost-energy counting
statistics, and yield the corresponding fluctuation
relations.

The framework of counting statistics can also constrain the precision of energy transfer. In particular, thermodynamic uncertainty relations express this connection by bounding the relative variance of currents or energy transfers in terms of entropy production, thereby limiting the precision attainable at a given thermodynamic cost
\cite{BaratoSeifert2015,
HorowitzGingrich2020, Strasberg2022}. For coherent perturbations, we derive a lower bound on the variance of the transferred boost energy in terms of its mean and the horizon temperature. This bound
follows from the exact counting statistics and
KMS spectral balance. Interestingly, Gallock-Yoshimura and Hasegawa~\cite{GallockHasegawa2026} obtained a similar bound by studying the stochastic entropy response of a black hole to infalling quantum radiation. Their analysis uses a scalar-field model with pure dephasing motivated by the Danielson--Satishchandran--Wald mechanism~\cite{Danielson2022, Danielson2023}, in which
photons carrying which-path information across the
horizon decohere a spatial superposition of an exterior
charge. Combining this description with
Einstein--Langevin dynamics, they obtain a lower bound
on the relative variance of the induced black-hole
entropy change in terms of the mean number of
entangling photons. Their result thus constrains a
stochastic gravitational response through photon
number.

Moreover, the relation between horizon entropy, energy transfer, and area variation naturally raises the question of whether the boost-energy fluctuation relations admit
a corresponding formulation for horizon-area changes.
This possibility is motivated by studies of black-hole
transitions in which gravitational backreaction relates
transition probabilities to area changes
\cite{MassarParentani2000,Iso2011}, and by the broader
statistical interpretation of causal horizons reviewed
by Jacobson and Parentani~\cite{JacobsonParentani2003}.
Here, we propose a stochastic horizon entropy--area
correspondence that associates each counted boost-energy
transfer with an area response,
extending the mean thermodynamic identification to
individual transfers. Under this additional hypothesis,
the boost-energy counting statistics determine the
area-change distribution, yielding integral and detailed
fluctuation relations and, for coherent perturbations,
an area-change uncertainty relation. The geometric interpretation of these results remains an assumption: a more fundamental justification would require deriving the area-change statistics from an independent quantum or stochastic description of the gravitational response
and establishing their agreement with the proposed
correspondence beyond the mean. Related quantum
approaches have obtained bounds and fluctuation formulas
for area observables associated with causal diamonds
\cite{CiambelliHeZurek2025,KowalskiGlikmanVarrin2026}.
These results offer complementary perspectives on
geometric fluctuations.

The manuscript is organized as follows.
Section~\ref{sec:ingredients} introduces the local
causal geometry, the horizon algebra, and the thermal
reference state, together with its modular structure.
Section~\ref{sec:results} develops the boost-energy
counting statistics, establishes the integral and
detailed fluctuation relations, and discusses their
extension to global bifurcate Killing horizons.
Section~\ref{sec:IV} specializes to coherent
perturbations, relates the mean transfer to the
boost-energy flux, and derives a thermodynamic
uncertainty relation.
Section~\ref{sec:area} reviews the mean geometric
response and introduces the additional stochastic
entropy--area correspondence used to obtain
area-change fluctuation relations and a corresponding
uncertainty bound for coherent perturbations.
Section~\ref{sec:conclusions} summarizes the results
and discusses their scope and possible extensions.
Throughout, we use $c=\hbar=k_B=1$ and the metric
signature $(-,+,+,+)$.

\section{Setting the stage}\label{sec:ingredients}
In this section, we review the geometric and algebraic ingredients needed to formulate energy-transfer statistics on a local causal horizon. We first introduce the local causal geometry and the boost flow associated with its future horizon. Following the construction of Dorau and Much, we then describe the horizon Weyl algebra and the quasifree reference state determined by the universal scaling-limit two-point function, together with its KMS property. Finally, we introduce the corresponding von Neumann algebra and use Tomita–Takesaki theory to identify the relation between the modular Hamiltonian and the boost generator. These ingredients provide the basis for the counting construction developed in the next section and are presented here to keep the discussion self-contained.

\subsection{Local causal horizons}\label{subsec:geometry}

\begin{figure}[t]
 \includegraphics[width=\columnwidth]{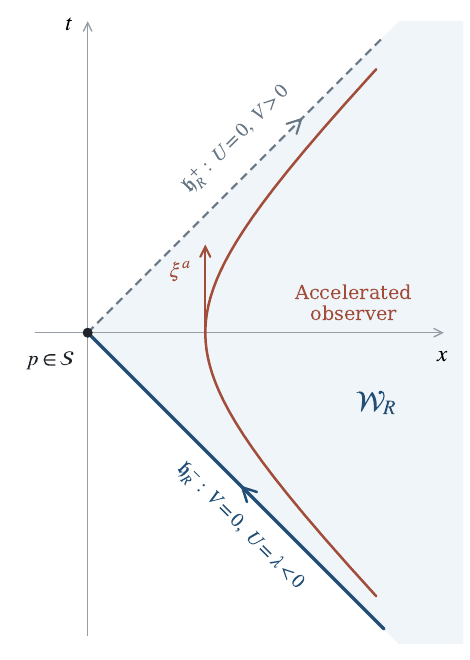}
 \caption{Local Rindler geometry, with transverse directions suppressed. The highlighted past horizon $\mathfrak{h}_R^-$ bounds the right wedge and is parametrized by $U = \lambda<0$. The point $p$ represents the chosen point on the cross section $\mathcal S$.}
 \label{fig:horizon}
\end{figure}

Let $p$ be a regular point of a smooth four-dimensional spacetime $(\mathcal{M},g_{ab})$. Choose a sufficiently small neighborhood $\mathcal{U}$ of $p$, with globally hyperbolic causal completion $\overline{\mathcal{U}}$, so that the metric in local inertial coordinates can be approximated by $g_{ab}|_\mathcal{U}\approx \eta_{ab}$, where $\eta_{ab}$ is the Minkowski metric. This approximation provides a local Minkowski
reference geometry. In a local inertial frame,
we choose a spatial axis and consider the Lorentz
boost flow in the plane spanned by this axis
and the time direction. Its generator $\xi^a$ is consequently an approximate Killing vector for the physical metric in $\mathcal{U}$. The associated past and future horizon branches, $\mathfrak{h}_R^-$ and $\mathfrak{h}_R^+$, meet at a spacelike two-surface $\mathcal S$ through the point $p$, as represented in Fig.~\ref{fig:horizon}. Along each branch, $\xi^a$ is null and both tangent and normal to the horizon, while it vanishes on $\mathcal S$. Together, these branches define the local causal horizon, an approximate bifurcate Killing horizon associated with the chosen boost flow~\cite{Guedens2012a, Guedens2012}.

In a neighborhood of $p$, we introduce double-null
coordinates $(U,V,x^i)$, with $i=1,2$, such that
$\mathcal S$ is given by $U=V=0$. The transverse
coordinates $x^i$ parametrize $\mathcal S$ and are
extended along the null generators, remaining
constant on each generator throughout the horizon
patch under consideration. Within the local Minkowski approximation, the metric takes the form
\begin{equation}
 ds^2=-\dd U\,\dd V+\delta_{ij}\dd x^i\dd x^j.\label{eq:metric}
\end{equation}
Choosing $\partial_U$ and $\partial_V$ to be future-directed, the right local wedge and its future and past local causal horizon branches are defined, respectively, by
\begin{equation}
\begin{aligned}
\mathcal W_R
 &= \bigl\{(U,V,x^1,x^2)\in\mathcal U
       \mid U<0,\ V>0\bigr\},\\
\mathfrak h_R^+
 &= \bigl\{(U,V,x^1,x^2)\in\mathcal U
       \mid U=0,\ V>0\bigr\}, \\
\mathfrak h_R^-
 &= \bigl\{(U,V,x^1,x^2)\in\mathcal U
       \mid U<0,\ V=0\bigr\}.
\end{aligned}\label{eq:lochor}
\end{equation}

We formulate our results on the past horizon $\mathfrak h_R^-$
bounding the right local wedge. Along $\mathfrak h_R^-$, we choose an
affine parameter $\lambda=U$ increasing towards the future, with
$\lambda=0$ at $\mathcal S$ and $\lambda<0$ on $\mathfrak h_R^-$.
The restriction of the boost Killing field to this branch is
\begin{equation}
 \xi^a=(\partial_v)^a
      =-\kappa\lambda(\partial_\lambda)^a,
 \label{eq:future}
\end{equation}
where $v=-\kappa^{-1}\log(\kappa\lambda)$ parametrizes the boost Killing flow. The constant $\kappa>0$ fixes the normalization of $\xi^a$ and hence the scale of boost energy. On $\mathfrak h_R^-$, it satisfies $\xi^a \nabla_a \xi^b=-\kappa\xi^b$.
For comparison, on the future branch $\mathfrak h_R^+$, an affine parameter $\lambda = V > 0$ increasing towards the future gives $\xi^a=\kappa\lambda(\partial_\lambda)^a$. On this branch, $\xi^a \nabla_a \xi^b=\kappa\xi^b$. Both expressions describe the same future-directed boost flow, whose orbits move away from $\mathcal S$ on $\mathfrak h_R^+$ and towards $\mathcal S$ on $\mathfrak h_R^-$.

The local Rindler description is used at leading order,
on scales for which curvature corrections to the locally
inertial metric, starting at quadratic order in the local
coordinates, remain negligible
\cite{Guedens2012a, Guedens2012}.
The causal completion of $\mathcal U$, denoted here by
$\overline{\mathcal U}$, is assumed to be globally
hyperbolic and to lie within the region where this
approximation remains valid, providing a suitable domain
for the local field dynamics.

Within this approximation, the horizon segment in
Eq.~\eqref{eq:lochor} is identified with a portion of
the associated Rindler reference half-horizon.
The horizon algebra is defined on this reference
half-horizon with $\lambda\in(-\infty,0)$, which is
preserved by dilations for every real parameter.
The exact boost evolution, modular structure, and
counting statistics employed below are defined in this
reference theory. For the application to the local
curved-spacetime setting, we consider perturbations
localized within the horizon segment contained in
$\mathcal U$, where the Rindler approximation is valid.


\subsection{Local horizon algebras}
\label{subsec:algebra}

Following the local construction of Dorau and Much \cite{DorauMuch2026}, we consider a real, minimally coupled free scalar field $\Phi$ of mass $m$ on the globally hyperbolic region $(\overline{\mathcal U},g_{ab})$. The field satisfies
\begin{equation}
  (\Box_g+m^2)\Phi=0,
  \qquad \Box_g:=-g^{ab}\nabla_a\nabla_b.
\end{equation}
To construct the algebra of observables on
$\mathfrak h_R^-$, we consider smooth, compactly
supported real characteristic data
$f,g\in C_0^\infty(\mathfrak h_R^-;\mathbb R)$
~\cite{Kay1991,Dappiaggi2009,Wall2012,KPV2021}.
We express these data in coordinates $(\lambda,x^i)$,
where $\lambda=U<0$. In these coordinates,
the symplectic form reads
\begin{equation}
  \sigma(f,g)
  =\int_{\mathfrak h_R^-}
  \bigl(f\,\partial_\lambda g-g\,\partial_\lambda f\bigr)
  \,\dd\lambda\,\dd\mathfrak a,
\end{equation}
with $\dd\mathfrak a$ denoting the transverse area
element on $\mathcal S$. The associated horizon Weyl algebra $\mathscr W_{\mathfrak h}$ is the unital
$C^*$-algebra generated by unitary operators $W(f)$
obeying
\begin{equation}
  W(f)^*=W(-f),\quad
  W(f)W(g)
  =e^{-\frac{i}{2}\sigma(f,g)}W(f+g).
  \label{eq:weyl}
\end{equation}
These relations encode the canonical commutation
relations in exponentiated form.

The Weyl algebra introduced above specifies the horizon observables and their commutation relations. To describe their fluctuations and thermal properties, we must also specify a reference state and its time evolution. Let $\tau_s$ denote the evolution induced by the
boost flow generated by $\xi^a$ in the reference theory. 
The geometric boost flow acts on the reference past
half-horizon as $\lambda\mapsto\ee^{-\kappa s}\lambda$.
Its action on characteristic data induces the automorphisms
\begin{equation}
  \tau_s\bigl(W(f)\bigr)=W(f_s),
  \qquad
  f_s(\lambda,x^i)=f(\ee^{+\kappa s}\lambda,x^i).
  \label{eq:horizon-boost-flow}
\end{equation}

Moreover, we take the bulk reference state to be centered, quasifree, invariant under this flow, and Hadamard across the reference horizon. Quasifreeness allows all correlation functions to be reconstructed from the two-point function, while the Hadamard condition fixes its local singularity structure. 
We use the universal scaling-limit covariance function associated with the horizon constructions of Refs.~\cite{Kay1991, KPV2021} to define
the reference horizon state. On the past half-horizon
$\mathfrak h_R^-$, it takes the form
\begin{equation}
  \Lambda(f,g)
  =-\frac{1}{\pi}\int_{\mathfrak{h}_R^{-}} \dd \mathfrak{a} \,
  \dd\lambda \, \dd\lambda'\,
  \frac{f(\lambda,x^i)\,g(\lambda',x^i)}
       {(\lambda-\lambda'-\ii0^+)^2},
  \label{eq:horizon-two-point}
\end{equation}
with the integral being understood distributionally. Its antisymmetric part reproduces the symplectic form, $2\operatorname{Im}\Lambda(f,g)=\sigma(f,g)$. The corresponding centered quasifree state $\omega_0$ on $\mathscr W_{\mathfrak{h}}$ is therefore specified by
\begin{equation}
  \omega_0\bigl(W(f)\bigr)
  =\exp\!\left[-\frac12\Lambda(f,f)\right].
  \label{eq:horizon-reference-state}
\end{equation}


The state $\omega_0$ satisfies the KMS condition~\cite{Haag1996} for evolution~\eqref{eq:horizon-boost-flow} at inverse temperature
\begin{equation}
  \beta_{\mathfrak{h}}=\frac{2\pi}{\kappa}.
\end{equation}
More explicitly, for $A,B\in\mathscr W_{\mathfrak{h}}$, the KMS function
\begin{align}
F_{A,B}(s)=\omega_0(A\tau_s(B)), \label{eq:KMSfunction}
\end{align}
admits a bounded analytic continuation to $0<\operatorname{Im}z<\beta_\mathfrak{h}$, continuous
on the closed strip, with boundary values
\begin{equation}
  F_{A,B}(s+\ii\beta_\mathfrak{h})
  =F_{B,A}(-s).
  \label{eq:horizon-KMS}
\end{equation}
This condition characterizes thermal equilibrium with respect
to the chosen evolution. 
Thus, the horizon state supplies the thermal reference for the subsequent energy-balance calculations. Its temperature is defined relative to boost Killing flow: rescaling $\xi^a\mapsto c\,\xi^a$, with $c>0$, gives $\kappa\mapsto c\kappa$ and $\beta_\mathfrak{h}\mapsto\beta_\mathfrak{h}/c$.

\subsection{Modular structure}

The centered quasifree horizon state $\omega_0$
is specified by $\Lambda$ through its expectation
values on Weyl operators. Its KMS property
expresses thermal equilibrium with respect
to the boost evolution $\tau_s$. To formulate the modular quantities needed below, we pass to the Gelfand--Naimark--Segal (GNS) representation~\cite{GelfandNaimark1943, Segal1947, Haag1996} of the horizon reference state
$\omega_0$. This provides the Hilbert-space setting
for constructing the horizon von Neumann algebra
and the associated modular operators.

Let $(\mathcal H_0,\pi_0,\Omega_0)$ be the GNS representation of $(\mathscr W_{\mathfrak{h}},\omega_0)$.
Here, $\mathcal H_0$ is a Hilbert space, while
$\pi_0:\mathscr W_{\mathfrak{h}}\to\mathcal B(\mathcal H_0)$ is a unital $*$-representation with $\mathcal B(\mathcal H_0)$ denoting the algebra of all bounded linear operators on $\mathcal H_0$. The normalized vector $\Omega_0\in\mathcal H_0$ represents the reference state $\omega_0$ through
\begin{equation}
  \omega_0(A)
  =\langle\Omega_0,\pi_0(A)\Omega_0\rangle,
  \qquad A\in\mathscr W_{\mathfrak{h}}.
\end{equation}
The associated horizon von Neumann algebra is
\begin{equation}
  \mathcal N_{\mathfrak{h}}
  :=\pi_0(\mathscr W_{\mathfrak{h}})'',
  \label{eq:horizon-von-neumann}
\end{equation}
where the prime denotes the commutant in
$\mathcal B(\mathcal H_0)$, namely the set of bounded operators commuting with every element of the indicated algebra. Henceforth, we suppress the representation map $\pi_0$ and write $A$ in place of $\pi_0(A)$.

The GNS vector $\Omega_0$ is cyclic and
separating for $\mathcal{N}_{\mathfrak{h}}$ \cite{KPV2021}, which allows us to apply Tomita--Takesaki modular theory~\cite{Haag1996, Takesaki1970, Borchers2000}.
The Tomita operator \(S\) is initially defined on the dense domain $\mathcal N_{\mathfrak h}\Omega_0$ by
\begin{equation}
  SA\Omega_0=A^\dagger\Omega_0,
  \qquad A\in\mathcal{N}_{\mathfrak{h}},
\end{equation}
and is understood henceforth as its closure.
Cyclicity makes this domain dense, while the separating
property makes the assignment well defined. Its polar decomposition is
\begin{equation}
  S=J\Delta_0^{1/2},
  \qquad
  \Delta_0=S^\dagger S,
  \qquad
  K=-\log\Delta_0.
  \label{eq:tomita}
\end{equation}
Here $J$ is the antiunitary modular conjugation,
with $J^2=\mathbb{I}$, and $\Delta_0$ is the positive,
self-adjoint, injective modular operator.
The self-adjoint operator $K$ is the dimensionless
modular Hamiltonian.


Then, the Tomita--Takesaki theorem gives
\begin{equation}
  J\mathcal{N}_{\mathfrak{h}}J=\mathcal{N}_{\mathfrak{h}}',
  \qquad
  \Delta_0^{\ii t}\mathcal{N}_{\mathfrak{h}}\Delta_0^{-\ii t}=\mathcal{N}_{\mathfrak{h}},
  \qquad t\in\mathbb R.
\end{equation}
Consequently, conjugation by $\Delta_0^{\ii t}$ defines
an automorphism group of the horizon von Neumann algebra.
The modular operators also satisfy
\begin{equation}
  J\Omega_0=\Omega_0,
  \qquad
  \Delta_0\Omega_0=\Omega_0,
  \qquad
  JK J=-K.
  \label{eq:modidentities}
\end{equation}

The KMS property now identifies the modular evolution
with the physical reference boost evolution. The invariance of $\omega_0$ under the action of $\tau_s$ allows the boost evolution to be implemented unitarily in its GNS representation. For the reference state considered here, this implementation is strongly continuous, so Stone's theorem yields a self-adjoint generator $L_\xi$ satisfying
\begin{equation}
  \ee^{\ii sL_\xi}A\Omega_0
  =\tau_s(A)\Omega_0,
  \qquad
  L_\xi\Omega_0=0. \label{eq:boostgenerator}
\end{equation}
Thus,
\begin{equation}
  \tau_s(A)
  =\ee^{\ii sL_\xi}A\ee^{-\ii sL_\xi}.
\end{equation}
With the KMS convention in Eq.~\eqref{eq:horizon-KMS},
uniqueness of the modular automorphism group gives
\begin{equation}
  \Delta_0^{\ii t}A\Delta_0^{-\ii t}
  =\tau_{-\beta_{\mathfrak{h}} t}(A).
  \label{eq:modular-boost-identification}
\end{equation}
Both implementations leave $\Omega_0$ invariant.
Their agreement on the dense set $\mathcal{N}_{\mathfrak{h}}\Omega_0$
therefore gives us
\begin{equation}
  \Delta_0=\ee^{-\beta_{\mathfrak{h}} L_\xi},
  \qquad
  K=\beta_{\mathfrak{h}} L_\xi.
  \label{eq:modular-energy}
\end{equation}
In particular, the modular action on past-horizon
Weyl operators is 
\begin{equation}
  \Delta_0^{\ii t}W(f)\Delta_0^{-\ii t}
  =W(f_{-\beta_{\mathfrak{h}}t}),
  \label{eq:modular-horizon-dilations}
\end{equation}
where $f_{-\beta_{\mathfrak{h}}t}(\lambda,x^i)
=f(\ee^{-2\pi t}\lambda,x^i).$ The negative exponent in the argument of $f$ follows from the inverse action of the geometric flow on characteristic data, evaluated at $s = - \beta_{\mathfrak{h}}t$. 

The geometric modular action in
Eq.~\eqref{eq:modular-boost-identification} is the local horizon counterpart of the Bisognano--Wichmann property~\cite{BisognanoWichmann1975} and is
closely related to the analysis of Summers and
Verch~\cite{SummersVerch1996}, who used modular inclusion to establish sufficient conditions fixing the temperature of equilibrium states to the Hawking value in spacetimes with a bifurcate Killing horizon. In the present setting, the connection between horizon dilations and thermal equilibrium is expressed by the relation
$K=\beta_{\mathfrak{h}}L_\xi$, which identifies the modular Hamiltonian with the rescaled boost generator of the reference horizon theory. In the next section, we use this identification to formulate the statistics of boost-energy transfer induced by unitary perturbations on the reference state.

\section{Modular counting statistics and fluctuation relations}
\label{sec:results}
We now formulate the statistics of boost-energy transfer induced by a unitary perturbation of the reference horizon state. We follow the algebraic approach to full counting statistics developed by Jak\v{s}i\'{c}, Pillet, and collaborators~\cite{JOPP2011,JOPS2012}, which relates statistics of energy exchange to the modular structure of the reference state. 

Let $\mathsf{U}\in\mathcal N_{\mathfrak h}$ be a unitary representing the
perturbation. The perturbation prepares
the state
\begin{equation}
 \omega_{\mathsf{U}}(A)
 =\omega_0(\mathsf{U}^\dagger A\mathsf{U})
 =\langle \mathsf{U}\Omega_0,A\,\mathsf{U}\Omega_0\rangle,
 \quad A\in\mathcal N_{\mathfrak h},
 \label{eq:perturbed}
\end{equation}
i.e., $\omega_{\mathsf{U}}$ is represented by $\psi_{\mathsf{U}} = \mathsf{U}\Omega_0$ in the GNS Hilbert space. Moreover, let $E_{L_\xi}$ denote the projection-valued spectral
measure of $L_\xi$. For every Borel set $B\subset\mathbb R$, define
\begin{equation}
 P_{\mathsf{U}}(B)
 :=\langle \mathsf{U}\Omega_0,E_{L_\xi}(B)\mathsf{U}\Omega_0\rangle
 =\|E_{L_\xi}(B)\mathsf{U}\Omega_0\|^2.
 \label{eq:counting-measure}
\end{equation}
The projection-valued nature of $E_{L_\xi}$ ensures positivity
and countable additivity, while unitarity gives $ P_{\mathsf{U}}(\mathbb R)=\|\mathsf{U}\Omega_0\|^2=1.
$ Hence $P_{\mathsf{U}}$ is a probability measure. We denote its real-valued counting variable by $q$. The corresponding dimensionless modular transfer is
\begin{equation}
 \Sigma=\beta_{\mathfrak h}q,
 \label{eq:modular-transfer}
\end{equation}
whose probability distribution is the spectral distribution of $K$ in $\mathsf{U}\Omega_0$. It represents the entropy-production variable in the sense of algebraic full counting statistics~\cite{JOPP2011,JOPS2012}.

By the spectral theorem, the characteristic function
of the boost-energy transfer $q$ is
\begin{align}
 \chi_{\mathsf{U}}(s)
 &:=\int_{\mathbb R}\ee^{\ii sq}P_{\mathsf{U}}(\dd q)
 \notag\\
 &=\langle \mathsf{U}\Omega_0,\ee^{\ii sL_\xi}\mathsf{U}\Omega_0\rangle
 \notag\\
 &=\langle\Omega_0,
       \mathsf{U}^\dagger\ee^{\ii sL_\xi}\mathsf{U}
       \ee^{-\ii sL_\xi}\Omega_0\rangle
 \notag\\
 &=\omega_0\bigl(\mathsf{U}^\dagger\tau_s(\mathsf{U})\bigr).
 \label{eq:counting}
\end{align}
Here we used the invariance of $\Omega_0$ and the unitary
implementation of $\tau_s$. The parameter $s$ is the Fourier variable conjugate to $q$ and enters the characteristic function through the reference boost evolution. The characteristic function uniquely determines the probability distribution. Whenever the corresponding moments are finite, they follow by differentiation at $s=0$, while derivatives of
$\log\chi_{\mathsf U}(s)$ give the cumulants. We use this representation in Sec.~\ref{sec:IV} to evaluate the coherent transfer statistics and obtain their mean and variance. Eq.~\eqref{eq:counting} expresses
$\chi_{\mathsf U}(s)$ as the expectation of
$\mathsf U^\dagger\tau_s(\mathsf U)
\in\mathcal N_{\mathfrak h}$ in the reference
state $\omega_0$. The counting statistics can
therefore be evaluated within the horizon algebra,
without introducing a reduced density matrix,
even though the spectral projections of $L_\xi$
used to define $P_{\mathsf U}$ need not themselves
belong to that algebra. 

Moreover, the standard generator $L_\xi$ accommodates positive and negative energy exchange. Indeed, $K=\beta_{\mathfrak h}L_\xi$ and $JKJ=-K$ imply $ JL_\xi J=-L_\xi,$ so its spectrum is symmetric about zero. This spectral symmetry allows transfer outcomes of either sign, but does not imply that $P_{\mathsf{U}}$ is symmetric, i.e.,  their probabilities depend on the perturbed vector $\mathsf{U}\Omega_0$. 

We now use the modular identities
established in the previous section to derive fluctuation relations for the probability measure~\eqref{eq:counting-measure}. It is worth mentioning that the following theorem applies the modular construction of Refs.~\cite{JOPP2011,JOPS2012} to perturbations of the horizon KMS reference state $\omega_0$ implemented by unitary operators $\mathsf U\in\mathcal N_{\mathfrak h}$, where the reference modular evolution is identified with the horizon boost flow. 

\begin{theorem}[Local-horizon boost-energy fluctuation relations]
\label{thm:local}
Let $\mathcal N_{\mathfrak h}$ be the von Neumann algebra
associated with the past horizon $\mathfrak h_R^-$ and
$\omega_0$ be its reference state satisfying the KMS condition at inverse temperature
$\beta_{\mathfrak h}=2\pi/\kappa$ with respect to the
boost evolution $\tau_s$. In the corresponding GNS representation, let $\Omega_0$ be the normalized cyclic and separating reference vector, and let $L_\xi$ be the standard self-adjoint boost generator given by Eq.~\eqref{eq:boostgenerator}. 

For any unitary $\mathsf{U}\in\mathcal N_{\mathfrak h}$, consider the perturbed state $\omega_{\mathsf{U}}(A)=\omega_0(\mathsf{U}^\dagger A\mathsf{U})$ with GNS representation $\psi_U = \mathsf U\,\Omega_0$ and its boost-energy probability measure defined by Eq.~\eqref{eq:counting-measure}. Then $P_{\mathsf{U}}$ is a probability measure satisfying
\begin{align}
 &\int_{\mathbb R}\ee^{-\beta_{\mathfrak h}q}P_{\mathsf{U}}(dq)
 =1,
 \label{eq:IFT}\\
 & P_{\mathsf{U}}(dq)
 =\ee^{\beta_{\mathfrak h}q}P_{\mathsf{U}^\dagger}(-dq).
 \label{eq:DFT}
\end{align}
Here $P_{\mathsf{U}^\dagger}$ is defined using the same reference
state and boost generator, with $\mathsf{U}$ replaced by
$\mathsf{U}^\dagger$, and $P_{\mathsf{U}^\dagger}(-dq)$ denotes its
reflection under $q\mapsto-q$.

Moreover, the mean energy transfer, $ \langle q\rangle_{\mathsf U}
 =\int_{\mathbb R}q\,P_{\mathsf U}(dq)$,
satisfies
\begin{equation}
 \beta_{\mathfrak h}\langle q\rangle_{\mathsf U}
 =S(\omega_{\mathsf U}\Vert\omega_0)\geq0,
 \label{eq:mean-D}
\end{equation}
where $S(\omega_{\mathsf U}\Vert\omega_0) =-\langle\psi_{\mathsf U},\log\Delta_{\Omega_0\mid\psi_{\mathsf U}}
\,\psi_{\mathsf U}\rangle$ is the Araki relative entropy between $\omega_{\mathsf{U}}$ and $\omega_0$~\cite{Araki1976}, with $\Delta_{\Omega_0\mid\psi_{\mathsf U}}$
being the relative modular operator. 
\end{theorem}

\begin{proof}

Because $\mathsf{U}\in\mathcal N_{\mathfrak h}$, the vector
$\mathsf{U} \Omega_0$ belongs to the domain of the Tomita operator,
and $S \mathsf{U} \Omega_0=\mathsf{U}^\dagger\Omega_0.$ The polar decomposition $S=J\Delta_0^{1/2}$ therefore
implies that $\mathsf{U} \Omega_0 \in\operatorname{Dom}\Delta_0^{1/2}$ and gives
\begin{equation}
 \mathsf{U}^\dagger\Omega_0
 =J\Delta_0^{1/2} \mathsf{U} \Omega_0
 =J\ee^{-\beta_{\mathfrak h}L_\xi/2}\mathsf{U} \Omega_0.
 \label{eq:tomita-step}
\end{equation}

The spectral theorem expresses its squared norm as
\begin{align}
   \left\|\ee^{-\beta_{\mathfrak h}L_\xi/2}\mathsf{U} \Omega_0 \right\|^2
 =\int_{\mathbb R}\ee^{-\beta_{\mathfrak h}q}P_{\mathsf{U}}(dq),   
\end{align}
and, since an antiunitary operator preserves norms, Eq.~\eqref{eq:tomita-step} yields
\begin{align}
 \int_{\mathbb R}\ee^{-\beta_{\mathfrak h}q}P_{\mathsf{U}}(dq)
 &=
 \left\|\ee^{-\beta_{\mathfrak h}L_\xi/2}\mathsf{U} \Omega_0 \right\|^2
 \notag\\
 &=
 \left\|J\ee^{-\beta_{\mathfrak h}L_\xi/2}\mathsf{U} \Omega_0 \right\|^2
 \notag\\
 &=\|\mathsf{U}^\dagger\Omega_0\|^2 \notag \\ & =
 1.
 \label{eq:IFTproof}
\end{align}
This proves the integral fluctuation relation given by Eq.~\eqref{eq:IFT}. 

To establish the detailed fluctuation relation, recall that
$JL_\xi J=-L_\xi$. Applying the spectral functional
calculus to the indicator function of a Borel set $B$
gives $JE_{L_\xi}(B)J=E_{L_\xi}(-B),
$
where $-B=\{-q:q\in B\}$. Consequently,
\begin{align}
 P_{\mathsf{U}^\dagger}(B)
 &=\|E_{L_\xi}(B)\mathsf{U}^\dagger\Omega_0\|^2
 \notag\\
 &=
 \left\|E_{L_\xi}(B)
 J\ee^{-\beta_{\mathfrak h}L_\xi/2}\mathsf{U} \Omega_0\right\|^2
 \notag\\
 &=
 \left\|JE_{L_\xi}(-B)
 \ee^{-\beta_{\mathfrak h}L_\xi/2}\mathsf{U} \Omega_0\right\|^2
 \notag\\
 &=
 \left\|E_{L_\xi}(-B)
 \ee^{-\beta_{\mathfrak h}L_\xi/2}\mathsf{U} \Omega_0\right\|^2
 \notag\\
 &=\int_{-B}\ee^{-\beta_{\mathfrak h}q}P_{\mathsf{U}}(dq).
 \label{eq:DFTproof}
\end{align}
The last equality follows because both operators are
functions of $L_\xi$ and their product multiplies the
spectral components of $\mathsf{U} \Omega_0$ by
$\mathbb{I}_{-B}(q)\ee^{-\beta_{\mathfrak h}q/2}$.
Replacing $B$ by $-B$ therefore gives
\begin{align}
   P_{\mathsf{U}^\dagger}(-B)
 =\int_B\ee^{-\beta_{\mathfrak h}q}P_{\mathsf{U}}(dq).  
\end{align}
Thus the reflected inverse-operation measure is
$\ee^{-\beta_{\mathfrak h}q}P_{\mathsf{U}}(dq)$, which proves
Eq.~\eqref{eq:DFT}.

It remains to establish Eq.~\eqref{eq:mean-D}. 
To evaluate the relative entropy between $\Omega_0$ and $\psi_{\mathsf{U}} = \mathsf{U} \Omega_0$, define the relative Tomita operator
on $\mathcal N_{\mathfrak h}\psi_{\mathsf{U}}$ by
\begin{equation}
 S_{\Omega_0\mid\psi_{\mathsf{U}}}(A\psi_{\mathsf{U}})=A^\dagger\Omega_0,
 \qquad A\in\mathcal N_{\mathfrak h},
 \label{eq:relative-tomita-definition}
\end{equation}
and denote its closure by the same symbol.
Because $\mathsf{U}$ is unitary,
$\mathcal N_{\mathfrak h}\psi_{\mathsf{U}} =\mathcal N_{\mathfrak h}\Omega_0$, i.e., $S$ and $S_{\Omega_0\mid\psi_{\mathsf{U}}}$ can therefore be compared on their common initial domain.
Writing $C=A\mathsf{U}$, so that $A=C\mathsf{U}^\dagger$, gives
\begin{align}
 S_{\Omega_0\mid\psi_{\mathsf{U}}}(C\Omega_0)
 =(C\mathsf{U}^\dagger)^\dagger\Omega_0
=\mathsf{U}C^\dagger\Omega_0
 =\mathsf{U}S(C\Omega_0).
 \label{eq:relative-tomita-action}
\end{align}
Since $\mathcal N_{\mathfrak h}\Omega_0$ is a core for
$S$ and $\mathsf{U}$ is bounded and unitary, this identity extends
to the closures.
Consequently, the corresponding relative modular
operator is
\begin{align}
 \Delta_{\Omega_0\mid\psi_{\mathsf{U}}}
 &=S_{\Omega_0\mid\psi_{\mathsf{U}}}^\dagger
   S_{\Omega_0\mid\psi_{\mathsf{U}}}=S^\dagger \mathsf{U}^\dagger \mathsf{U}S
 =S^\dagger S =\Delta_0.
 \label{eq:relative-modular-unitary}
\end{align}
Araki's formula therefore gives
\begin{align}
 S(\omega_{\mathsf{U}}\Vert\omega_0)
 &=-\langle\psi_{\mathsf{U}},
       \log\Delta_{\Omega_0\mid\psi_{\mathsf{U}}}\,\psi_{\mathsf{U}}\rangle
 \notag\\
 &=-\langle\psi_{\mathsf{U}},\log\Delta_0\,\psi_{\mathsf{U}}\rangle
 \notag\\
 &=\beta_{\mathfrak h}
   \int_{\mathbb R}qP_{\mathsf{U}}(dq)
 \notag\\
 &=\beta_{\mathfrak h}\langle q\rangle_{\mathsf{U}}. 
 \label{eq:mean-D-proof}
\end{align}

\end{proof}

First, the fluctuation relations are formulated directly in terms of probability measures and do not require
probability densities. Here, $P_{\mathsf U}(dq)$
denotes the measure over possible boost-energy
transfers. When densities exist, this notation becomes
$P_{\mathsf U}(dq)=p_{\mathsf U}(q)\,dq$ and
$P_{\mathsf U^\dagger}(-dq)=p_{\mathsf U^\dagger}(-q)\,dq$.
Within the local horizon reference theory,
Eq.~\eqref{eq:IFT} is the Jarzynski
equality~\cite{Jarzynski1997} for boost-energy transfer,
while Eq.~\eqref{eq:DFT} is the corresponding Crooks
detailed fluctuation relation~\cite{Crooks1998},
comparing a unitary operation with its
inverse~\cite{JOPP2011,JOPS2012}. Integrating this
detailed relation with the appropriate exponential
weight yields the integral identity.

With energy counted relative to the boost field
$\xi^a$ and thermal weighting fixed by
$\beta_{\mathfrak h}=2\pi/\kappa$,
Eq.~\eqref{eq:DFT} relates a transfer $q$ under
$\mathsf U$ to the opposite transfer under
$\mathsf U^\dagger$ through the statistical factor
$\ee^{\beta_{\mathfrak h}q}$. This comparison involves
two separate preparations that begin in $\omega_0$
and use the same horizon algebra and boost evolution,
with $\mathsf U$ applied in the forward preparation
and $\mathsf U^\dagger$ in the reverse one.
In particular, the reverse preparation produces
$\omega_{\mathsf U^\dagger}(A)
=\omega_0(\mathsf U A\mathsf U^\dagger)$.


Finally, Eq.~\eqref{eq:mean-D} identifies the mean entropy production
with the relative entropy of the perturbed state, i.e., $
 \langle\Sigma\rangle_{\mathsf{U}}
 =\beta_{\mathfrak h}\langle q\rangle_{\mathsf{U}}
 =S(\omega_{\mathsf{U}}\Vert\omega_0)\geq0.$
This gives a second-law statement for the modular counting
statistics. The mean entropy production vanishes precisely when $\omega_{\mathsf{U}}=\omega_0$, that is, when the operation leaves the reference state invariant.

\subsection{Extension to global bifurcate Killing horizons}
\label{subsec:other}

The preceding analysis could be extended from local
causal horizons to spacetimes possessing a global bifurcate Killing horizon, by adapting the construction developed in Ref.~\cite{Kay1991}. In this setting, the horizon evolution is generated by an exact Killing symmetry of the background spacetime. Then,
consider a real, minimally coupled free scalar field admitting a centered, Killing-invariant quasifree reference state that is regular across the horizon. Under the hypotheses of the Kay--Wald thermal
result, the state induced on the relevant half-horizon algebra is KMS with respect to the horizon Killing evolution, at inverse temperature $\beta_{\mathfrak{h}}=2\pi/\kappa$.

The construction then follows the same steps as in
Sec.~\ref{sec:ingredients}, using the corresponding horizon algebra, reference state, and Killing evolution. 
Consequently, the proof of Theorem~\ref{thm:local} would be applied directly, and every unitary $\mathsf{U}\in\mathcal{N}_{\mathfrak{h}}$ obeys Eqs.~\eqref{eq:IFT}--\eqref{eq:mean-D}. We do not explicitly make this extension here.

For the maximally extended Schwarzschild spacetime, the
Hartle--Hawking--Israel state provides a natural candidate for the reference state. Its Hadamard property across the bifurcate horizon ensures that it satisfies the regularity assumption above. Through the horizon-algebra construction described in Sec.~\ref{sec:ingredients}, it induces a state on the chosen past half-horizon algebra that satisfies the KMS condition at inverse Hawking temperature $\beta_{\mathfrak{h}}=8\pi GM$ with respect to the induced Schwarzschild Killing evolution~\cite{Sanders15}, where $G$ is the gravitational constant and $M$ is the mass of the black hole.

\section{Coherent perturbations and thermodynamic uncertainty}
\label{sec:IV}

In this section, we specialize the counting construction to coherent scalar-field perturbations. We compute their characteristic function and first two cumulants, and identify the mean transfer with the excess boost-energy flux through $\mathfrak h_R^-$. Finally, we use the KMS condition to derive a thermodynamic uncertainty relation that bounds the variance of the boost-energy transfer in terms of its mean transfer.

\subsection{Coherent perturbations and boost-energy transfer}\label{subsec:coherent}

Consider a coherent perturbation with profile
$f\in C_0^\infty(\mathfrak h_R^-;\mathbb R)$.
The Weyl operation $\mathsf{U}=W(f)$ transforms the reference state
$\omega_0$ into $\omega_f(A)=\omega_0\!\left(W(f)^* A W(f)\right).$ We denote the characteristic function of the resulting boost-energy transfer by $ \chi_f(s)=\chi_{W(f)}(s)$. Using $\tau_s(W(f))=W(f_s)$, Eqs.~\eqref{eq:weyl}, ~\eqref{eq:horizon-reference-state}, and~\eqref{eq:counting} give us
\begin{align}
 \log\chi_f(s)
 &=\frac{\ii}{2}\sigma(f,f_s)-\frac12\Lambda(f_s-f,f_s-f),
 \label{eq:coherent-CF}
\end{align}
where $ f_s(\lambda,x^i)=f(\ee^{+\kappa s}\lambda,x^i)$. Eq.~\eqref{eq:coherent-CF} gives an exact
closed-form expression for the characteristic
function, entirely in terms of the horizon
symplectic form $\sigma$ and the covariance function $\Lambda$, with the dependence
on $s$ entering through the boost-transformed
data $f_s$. 

The covariance term in Eq.~\eqref{eq:coherent-CF}
has vanishing first derivative at $s=0$, since
$f_s-f=0$ there. Consequently, the mean transfer is
\begin{align}
 \langle q\rangle_f
 &=-\ii\left.\frac{d}{ds}\log\chi_f(s)\right|_{s=0} = \frac12\sigma(f,\mathcal D_\xi f) \nonumber \\ &
=-\kappa\int_{\mathfrak{h}_R^{+}}d{\mathfrak{a}}\,
        d\lambda\,\lambda(\partial_\lambda f)^2,
 \label{eq:coherent-mean}
\end{align}
where $\mathcal D_\xi f
:=\left.\frac{d}{ds}f_s\right|_{s=0}
=\kappa \lambda\partial_\lambda f$. The last equality follows by integration by parts, with all boundary terms vanishing because of compact support. Now, to identify the mean transfer with the coherent excess energy flux, let
$:T_{ab}:$ denote the minimally coupled scalar
stress--energy tensor normal ordered with respect to
the reference Hadamard state $\omega_0$. This is defined
by point splitting with subtraction of the reference
two-point function, so that
$\langle:T_{ab}:\rangle_{\omega_0}=0$. For horizon data $f$ induced by a smooth real Klein--Gordon solution, the corresponding bulk coherent state $\omega_f$ satisfies
\begin{equation}
 \langle:T_{ab}:\rangle_{\omega_f}k^ak^b
 =\langle:T_{\lambda \lambda}:\rangle_{\omega_f}
 =(\partial_\lambda f)^2,
 \label{eq:coherent-null-stress}
\end{equation}
where $k^a=(\partial_\lambda)^a$. Using $\xi^a=-\kappa \lambda k^a$
on $\mathfrak h_R^-$, Eq.~\eqref{eq:coherent-mean}
therefore becomes
\begin{align}
 \langle q\rangle_f =\int_{\mathfrak h_R^-}
   \langle:T_{ab}:\rangle_{\omega_f} \xi^a k^b\,d\lambda\,d\mathfrak{a}.
 \label{eq:flux}
\end{align}
Thus, the first cumulant is the boost-energy flux
through the past horizon, measured relative to
$\omega_0$. This is the coherent flux relation employed by
Dorau and Much~\cite{DorauMuch2026}. Related expressions for coherent-state relative entropy in Minkowski wedges were obtained in Refs.~\cite{Longo2019, CasiniGrilloPontello2019}.

Invariance of the symplectic form under the boost flow implies
\begin{align}
  \sigma(f,\mathcal D_\xi^2f)=-\sigma(\mathcal D_\xi f,\mathcal D_\xi f)=0.
\end{align}
The symplectic term therefore does not contribute to
the second cumulant, and
\begin{equation}
 \operatorname{Var}_f(q)
 =-\left.\frac{d^2}{ds^2}\log\chi_f(s)\right|_{s=0}
 =\Lambda(\mathcal D_\xi f,\mathcal D_\xi f),
 \label{eq:coherent-variance}
\end{equation}
where second cumulant measures the fluctuations of the
boost-energy transfer around its mean, i.e., $\langle q \rangle_f$, $\operatorname{Var}_f(q)=\langle(q-\langle q \rangle_f)^2\rangle_f$.
Since $\mathcal D_\xi f$ is smooth and compactly
supported, $\Lambda(\mathcal D_\xi f,\mathcal D_\xi f)$
is finite, and hence so is the transfer variance. 
Its expression in terms of $\Lambda$ shows that these
fluctuations are determined by the reference-state
covariance two-point function evaluated on $\mathcal D_\xi \, f$. Thus, even for a precisely
specified profile $f$, the transferred energy generally
has a nonzero variance.


\subsection{A coherent thermodynamic uncertainty relation}
\label{subsec:TUR}

The preceding subsection expressed the mean boost-energy transfer as the classical excess flux of the coherent perturbation and its variance in terms of the covariance function. We now use the KMS condition to relate these quantities and derive a lower bound on the variance for a given mean transfer. 

By defining $ A_f=\Phi(f)$  and $B_f=\Phi(\mathcal D_\xi f)$, where $\Phi(f)$ is the horizon field generator defined by $W(tf)=\ee^{\ii t\Phi(f)}$, with
$\omega_0(\Phi(f)\Phi(g))=\Lambda(f,g)$,
we can express Eqs.~\eqref{eq:coherent-mean} and~\eqref{eq:coherent-variance} as
\begin{equation}
2\langle q\rangle_f = -\ii\,\omega_0([A_f,B_f]), \quad \operatorname{Var}_f(q) = \omega_0(B_f^2).
 \label{eq:coherent-moments-fields}
\end{equation}
The KMS function given by Eq.~\eqref{eq:KMSfunction} is useful here because its two boundary values contain the two operator orderings needed to form the commutator. Specifically,
\begin{align}
  F_{A_f,B_f}(0)=\omega_0(A_fB_f),
 \quad
 F_{A_f,B_f}(\ii\beta_{\mathfrak{h}})=\omega_0(B_fA_f), \nonumber  
\end{align}
which allow us to obtain
\begin{align}
 \ii\bigl[
 F_{A_f,B_f}(\ii\beta_h)-F_{A_f,B_f}(0)
 \bigr]
 & =-\ii\,\omega_0([A_f,B_f]) \nonumber \\ &
 =2\langle q\rangle_f.
 \label{eq:mean-KMS-boundaries}
\end{align}
Eq.~\eqref{eq:mean-KMS-boundaries} expresses the mean
through a difference of boundary values. To connect it
to the variance, we convert that difference into an
integral of the KMS function of $B_f$ with itself. In order to do this, we can rewrite Eq.~\eqref{eq:KMSfunction} as $ F_{A_f,B_f}(s)
 =\omega_0(\tau_{-s}(A_f)B_f).$ This form is useful because differentiating $A_f$ replaces
it by $B_f$, i.e.,
\begin{align}
 \frac{d}{ds}F_{A_f,B_f}(s)
 &=-\omega_0(\tau_{-s}(B_f)B_f)
 \notag\\
 &=-F_{B_f,B_f}(s).
\end{align}
Substituting this derivative evaluated on the segment
$z=\ii u$ into 
Eq.~\eqref{eq:mean-KMS-boundaries} yields
\begin{align}
 2\langle q\rangle_f
 &=\ii\int_0^{\beta_h}du\,
   \frac{d}{du}F_{A_f,B_f}(\ii u)
 \notag\\
 &=\int_0^{\beta_h}du\,F_{B_f,B_f}(\ii u).
 \label{eq:coherent-KMS-sum-rule}
\end{align}

It remains to compare the integrand
$F_{B_f,B_f}(\ii u)$ with
$F_{B_f,B_f}(0)=\operatorname{Var}_f(q)$
throughout $0\leq u\leq\beta_{\mathfrak h}$. First, notice that
\begin{align}
F_{B_f,B_f}(s)
=\omega_0\!\left(B_f\tau_s(B_f)\right)
=\left\langle B_f\Omega_0,e^{isL_\xi}B_f\Omega_0\right\rangle, \nonumber
\end{align}
which gives
\begin{equation}
 F_{B_f,B_f}(\ii u)
 =\left\|\ee^{-uL_\xi/2}B_f\Omega_0\right\|^2. \label{eq:Fiu}
\end{equation}
Eq.~\eqref{eq:Fiu} establishes positivity and convexity
in $u$. 
Given that a convex function lies below the straight line joining its endpoints and the KMS condition also fixes both endpoint values, i.e., $ F_{B_f,B_f}(0)
 =F_{B_f,B_f}(\ii\beta_h)
 =\omega_0(B_f^2)$, we obtain
\begin{equation}
 0\leq F_{B_f,B_f}(\ii u)
 \leq\omega_0(B_f^2)
 =\operatorname{Var}_f(q).
 \label{eq:KMS-covariance-bound}
\end{equation}
Integrating Eq.~\eqref{eq:KMS-covariance-bound} over the
KMS interval and using
Eq.~\eqref{eq:coherent-KMS-sum-rule} give us $ \operatorname{Var}_f(q)
 \geq\frac{2}{\beta_\mathfrak{h}}\langle q\rangle_f$, i.e.,
\begin{equation}
 \frac{\operatorname{Var}_f(q)}{\langle q\rangle_f^2}
 \geq\frac{2}{\beta_h\langle q\rangle_f}.
\label{eq:energy-TUR}
\end{equation}

Eq.~\eqref{eq:energy-TUR} is a thermodynamic uncertainty relation for coherent boost-energy transfer, expressing a trade-off between precision and thermodynamic
cost~\cite{BaratoSeifert2015, HorowitzGingrich2020}, i.e., at fixed temperature and positive mean transfer, the variance cannot be arbitrarily small. When $0<\langle q\rangle_f\ll T_{\mathfrak{h}}=\beta_\mathfrak{h}^{-1}$, this cost is small and the energy transfer necessarily exhibits large fluctuations relative to its mean. Increasing $\beta_h\langle q\rangle_f$ lowers the bound on relative fluctuations. Moreover, using $\Sigma=\beta_\mathfrak{h}q$ and $S(\omega_f\Vert\omega_0)=\beta_\mathfrak{h}\langle q\rangle_f$,
the same bound becomes
$\operatorname{Var}_f(\Sigma)\geq2S(\omega_f\Vert\omega_0)$.
This dimensionless formulation is invariant under a consistent positive rescaling of the Killing generator. Indeed, $\xi^a\mapsto c\,\xi^a$, with $c>0$, induces $q\mapsto cq$ and $\beta_\mathfrak{h}\mapsto\beta_\mathfrak{h}/c$, leaving $\Sigma$ and the cost $\beta_\mathfrak{h}\langle q\rangle_f$ unchanged.


\section{A stochastic horizon-area response?}
\label{sec:area}

A natural next step is to ask whether the boost-energy
fluctuation relations, Eqs.~\eqref{eq:IFT} and
\eqref{eq:DFT}, admit a corresponding formulation for
horizon-area changes. Such a formulation requires a
probability distribution for the area variation. The modular counting law $P_{\mathsf{U}}$ does not, by itself, supply this geometric information. Nevertheless, for coherent perturbations, the mean energy transfer already has a gravitational interpretation through its relation to
the horizon-area response~\cite{Jacobson1995, DorauMuch2026}. This mean relation motivates exploring a stochastic extension of the horizon entropy--area correspondence.

To explore this possibility, we postulate that each
counted boost-energy transfer determines a geometric
area change through the same proportionality that
relates their means. Once this identification is assumed, the area-change distribution is fixed by a rescaling of the modular counting law, and the corresponding fluctuation
relations follow directly. 

A more fundamental justification would require
obtaining area-change statistics from an independent
gravitational description, for example a quantum or
stochastic model of the horizon response. These
statistics could then be compared with those inferred
from modular boost-energy counting, using the relations
derived here as tests of the proposed correspondence.

\subsection{Mean area response}
\label{subsec:mean-area}

As discussed in Sec.~\ref{sec:IV}, for a coherent perturbation $\mathsf{U}=W(f)$, Eq.~\eqref{eq:flux} identifies the mean transfer with the excess boost-energy flux. Together with Eq.~\eqref{eq:mean-D}, this yields
$S(\omega_f\Vert\omega_0)
=\beta_{\mathfrak h}\langle q\rangle_f,$ reproducing the relation obtained by Dorau and Much~\cite{DorauMuch2026}. To connect this field-theoretic result to the geometric response of the horizon, the authors introduce the additional assumption that the relative entropy is proportional to a first-order variation of the horizon area, implementing this proportionality through an ansatz for the induced transverse metric. Comparison with
the leading-order Raychaudhuri equation about an
equilibrium horizon, together with stress-energy
conservation, yields the semiclassical Einstein field
equations. Fixing the proportionality to the
Bekenstein--Hawking normalization gives
\begin{equation}
 \frac{\delta\mathfrak {a}}{4G} = \beta_{\mathfrak h}\langle q\rangle_f
 =  S(\omega_f\Vert\omega_0),
 \label{eq:mean-area-response}
\end{equation}
where $\delta\mathfrak a
 =\int_{\mathfrak{h}_R^-}\theta\,d\lambda\,d\mathfrak{a}$, with $\theta$ being
fractional rate of change of the transverse area
element.

The thermodynamic content of this identification
can be traced back to Jacobson's
formulation~\cite{Jacobson1995}. His heat
$\delta Q$ is the matter boost-energy flux across
the local causal horizon, $\delta Q = \int_{\mathfrak h_R^{-}} T_{ab}\xi^a\,d\mathfrak{a}^b,$ with $\delta Q$ representing a macroscopic energy transfer. Moreover, the local causal horizon is chosen to have vanishing expansion and shear at the equilibrium cross section and its temperature is the Unruh temperature associated with the approximate boost flow. The two thermodynamic assumptions underlying the derivation
of the Einstein field equations are the Clausius relation
and the horizon entropy--area relation,
\begin{equation}
 \delta Q=T_{\mathfrak h}\,\delta S_{\mathfrak h},
 \qquad
 \delta S_{\mathfrak h}=\eta\,\delta\mathfrak a,
 \label{eq:jacobson-assumptions}
\end{equation}
where $\eta=1/(4G)$ for Einstein gravity. Together, these
assumptions relate the heat flux $\delta Q$ through the horizon to the corresponding area change $\delta\mathfrak a$ through $\beta_{\mathfrak h}\delta Q=\delta\mathfrak a/(4G)$. 




\subsection{A stochastic horizon entropy–area correspondence}
\label{subsec:stochastic-clausius}

Let us start by recalling that the relation between heat and entropy along individual trajectories are already part of stochastic thermodynamics~\cite{Strasberg2022}. 
For a reservoir maintained at inverse temperature $\beta$,
the heat $Q[\gamma]$ absorbed along a stochastic trajectory $\gamma$ contributes $+\beta Q[\gamma]$ to the reservoir entropy~\cite{Seifert2012}. The trajectory $\gamma$ describes the history of the system during a single realization of a specified protocol, as represented by the sequence of states visited or, in a monitored quantum process, by the measurement
record. The heat $Q[\gamma]$ is thus assigned to that particular history and can fluctuate between realizations of the same protocol.

Black-hole transition processes also provide motivation for extending the mean horizon entropy–area relation to individual realizations. Massar and Parentani~\cite{MassarParentani2000} incorporate
the geometric response to Hawking emission at the
level of transition amplitudes and find that
emission probabilities are governed by the
associated horizon-area change. Building on this
result, Iso, Okazawa, and Zhang~\cite{Iso2011}
use a transition-probability ratio of the form
\begin{equation}
 \frac{ \mathcal W(\mathcal C\rightarrow\mathcal C')}
 { \mathcal W(\mathcal C'\rightarrow\mathcal C)}
 =  \exp\!\left(
 \frac{\mathfrak a'-\mathfrak a}{4G}
 \right),
 \label{eq:black-hole-area-balance}
\end{equation}
where $\mathcal C$ and $\mathcal C'$ are
configurations with horizon areas $\mathfrak a$
and $\mathfrak a'$. They subsequently consider
sequences of such transitions, making the net
area change a quantity associated with an
individual stochastic history. Their fluctuation relation also includes the initial canonical distributions of the matter outside the horizon for the forward
and reverse protocols. These distributions contribute the term $\beta_{\mathfrak h}(\Delta E-\Delta F)$, where $\Delta E$ and $\Delta F$
denote the energy and equilibrium free-energy differences of matter, in addition to the change in black-hole entropy.

Moreover, this motivation extends beyond black holes.
In their review on horizon thermodynamics, Jacobson and Parentani~\cite{JacobsonParentani2003} argue that the
statistical interpretation of horizon entropy extends
beyond black holes to causal horizons more generally.
In this picture, the familiar geometric properties
describe ensemble averages, while energy exchanges
can produce fluctuations of the horizon area. This perspective motivates a statistical description of the local causal horizon response. The additional step proposed here is to relate its area changes to individual counted boost-energy transfers.

For a fixed perturbation $\mathsf U$, let $q$ denote
the boost-energy counting variable with probability
measure $P_{\mathsf U}$ defined by
Eq.~\eqref{eq:counting-measure}. We suppose that the
horizon response admits a description in which each
counted transfer $q$ determines a area change
$\delta\mathfrak a(q)$ between prescribed initial and
final cuts, evaluated to first order in the metric
perturbation. Then, we identify the dimensionless modular transfer $\Sigma(q)=\beta_{\mathfrak h}q$, defined in
Eq.~\eqref{eq:modular-transfer}, with the horizon
entropy change $\delta S_{\mathfrak h}(q)$.
We further postulate the stochastic horizon
entropy--area correspondence:
$\delta S_{\mathfrak h}(q)=\delta\mathfrak a(q)/(4G)$.
Combining these identifications gives
\begin{equation}
 \frac{\delta\mathfrak a(q)}{4G}
 =\beta_{\mathfrak h}q
 =\Sigma(q).
 \label{eq:jacobsonstoch}
\end{equation}
The sign convention assigns a
positive area change to a positive counted
boost-energy transfer. At fixed $q$, the area
response is determined by this correspondence and
therefore its fluctuations are inherited 
from the counting statistics of $q$. Therefore, Eq.~\eqref{eq:jacobsonstoch} strengthens
the thermodynamic assumptions in Eq.~\eqref{eq:jacobson-assumptions} by imposing the entropy--area identification at the level of individual transfers.

The corresponding area-change probability measure
is denoted by $P_{\mathsf U}^{\mathfrak a}$, which corresponds to the rescaled image of $P_{\mathsf{U}}$,  i.e., 
\begin{equation}
 P_{\mathsf{U}}^{\mathfrak{a}}(B_{\mathfrak{a}}) =P_{\mathsf{U}}\bigl(B\bigr),
 \label{eq:area-pushforward}
\end{equation}
with $B_{\mathfrak{a}}= \bigl\{4G\beta_{\mathfrak h}q:\ q\in B\bigr\}$. Moreover, the full area-change statistics are fixed by the boost-energy statistics once
Eq.~\eqref{eq:jacobsonstoch} is assumed. Combining Eqs.~\eqref{eq:jacobsonstoch},~\eqref{eq:area-pushforward},  and ~\eqref{eq:mean-D} yields
\begin{equation}
 \frac{\langle\delta\mathfrak a\rangle_{\mathsf U}}{4G}
 =\beta_{\mathfrak h}\langle q\rangle_{\mathsf U}
 =S(\omega_{\mathsf U}\Vert\omega_0)\geq0,
 \label{eq:stochastic-area-mean}
\end{equation}
which expresses the second law for the local causal
horizon on average. For coherent perturbations, the mean area change reproduces the macroscopic geometric response
in Eq.~\eqref{eq:mean-area-response}, as expected. It is worth mentioning that the nonnegative mean area response concerns unitary perturbations of the specified equilibrium reference. For an eternal Schwarzschild black hole, this reference is supplied by the Hartle--Hawking--Israel state, which does not describe an evaporating black hole.



Expressing the integral fluctuation relation~\eqref{eq:IFT} in terms of area change implies
\begin{align}
 \left\langle
 \ee^{-\delta\mathfrak a/(4G)}
 \right\rangle_{\mathsf{U}}
  =1.
 \label{eq:area-IFT}
\end{align}
This is the corresponding Jarzynski
identity for the area response. Likewise, expressing both energy-transfer distributions
in Eq.~\eqref{eq:DFT} in terms of the corresponding area change gives
\begin{equation}
 P_{\mathsf{U}}^{\mathfrak a}(d\,\delta\mathfrak a)
 =e^{\delta\mathfrak a/(4G)}
 P_{\mathsf{U}^\dagger}^{\mathfrak a}(-d\,\delta\mathfrak a).
 \label{eq:area-detailed-fluctuation}
\end{equation}
Here, $P_{\mathsf U}^{\mathfrak a}(d\,\delta\mathfrak a)$
denotes the probability measure over the possible values
of the area change $\delta\mathfrak a$. The notation
$d\,\delta\mathfrak a=d(\delta\mathfrak a)$ indicates
integration with respect to this variable and does not
represent a second variation of the area. The inverse-operation measure $P_{\mathsf{U}^\dagger}^{\mathfrak a}$
is defined as in Eq.~\eqref{eq:area-pushforward},
with $\mathsf{U}$ replaced by $\mathsf{U}^\dagger$ and the same reference state $\omega_0$. The notation
$P_{\mathsf{U}^\dagger}^{\mathfrak a}(-d\,\delta\mathfrak a)$ denotes its reflection under
$\delta\mathfrak a\mapsto-\delta\mathfrak a$. 

The integral fluctuation relation~\eqref{eq:area-IFT}
implies that a positive mean area increase must be
accompanied by a nonzero probability of area decrease.
It also places an exponential upper bound on the
probability of decreases exceeding a given magnitude:
for any threshold $a_*>0$, $ P_{\mathsf U}^{\mathfrak a}\bigl((-\infty,-a_*]\bigr)\leq e^{-a_*/(4G)}.$ Thus, the probability of an area decrease of magnitude at least $a_*$ is exponentially bounded by that magnitude in units of $4G$. The detailed relation~\eqref{eq:area-detailed-fluctuation} further relates an area increase $\delta\mathfrak a>0$ under $\mathsf U$ to a decrease of the same magnitude under $\mathsf U^\dagger$, with relative statistical weight $e^{\delta\mathfrak a/(4G)}$ favoring the increase.

In addition, for coherent perturbations, the thermodynamic uncertainty relation~\eqref{eq:energy-TUR} gives an area-change uncertainty relation
\begin{equation}
 \operatorname{Var}_f(\delta \mathfrak{a})\geq8G\langle\delta \mathfrak{a}\rangle_f.
 \label{eq:area-uncertainty}
\end{equation}
Thus, for a fixed positive mean increase in horizon area,
the fluctuations have a nonzero lower bound. Related studies have investigated area fluctuations
in finite causal diamonds in Minkowski spacetime. The authors of Ref.~\cite{CiambelliHeZurek2025} quantize the constrained gravitational phase space of a stretched horizon and derive a lower bound on the area variance in the null limit. Building on this phase-space construction, the authors of Ref.~\cite{KowalskiGlikmanVarrin2026} quantize a subalgebra of time-averaged geometric variables
and construct a thermal state. In the null limit,
the resulting area variance contains contributions
that are linear and quadratic in the mean area. On the other hand, inequality~\eqref{eq:area-uncertainty} concerns the area change of a local
causal horizon induced by a prescribed coherent
perturbation. It follows from boost-energy counting
statistics together with the assumptions in
Eqs.~\eqref{eq:jacobsonstoch} and~\eqref{eq:area-pushforward}.

\section{Conclusions}
\label{sec:conclusions}

We have formulated boost-energy-transfer statistics
for local causal horizons within the algebraic
framework of quantum field theory. The thermal
reference state identifies the modular generator
with the generator of boost evolution, allowing
the established modular counting construction
to be given a horizon-energy interpretation. The resulting probability distribution satisfies integral and detailed fluctuation relations, with the latter comparing a unitary perturbation $\mathsf U$ and its inverse $\mathsf U^\dagger$ applied separately to the same reference state.
The mean dimensionless transfer equals Araki relative entropy, providing the corresponding second-law statement.
These results remain meaningful for type-III algebras
without introducing a reduced density matrix.
The construction could, in principle, be extended to global bifurcate Killing horizons admitting the required thermal reference state.

For coherent scalar-field perturbations, we obtained
the exact characteristic function and its first two
cumulants. The mean transfer is the excess boost-energy
flux through the horizon, recovering the field-theoretic
relation used by Dorau and Much~\cite{DorauMuch2026},
while the variance is determined by the scaling-limit covariance function. Positivity and convexity of
the imaginary-time KMS correlation function then yield
a thermodynamic uncertainty relation.
This bound limits the precision of coherent energy
transfer at a given positive mean and horizon
temperature. An open question is whether a thermodynamic uncertainty relation can be established for arbitrary unitary perturbations $U\in\mathcal N_{\mathfrak h}$.

We subsequently examined whether these energy
statistics admit a geometric interpretation, starting from the relation between mean energy transfer and horizon-area response underlying the approaches of Jacobson and Dorau and Much~\cite{Jacobson1995,DorauMuch2026}.
To extend this relation beyond the mean, we postulated
a stochastic correspondence associating each counted
boost-energy transfer with a horizon-area change.
Under this stochastic horizon entropy--area
correspondence, the energy statistics yield integral
and detailed relations for area changes, together
with the coherent uncertainty bound. However, explaining the dynamics underlying these area
fluctuations requires a gravitational description
beyond the modular construction. 


A direct next step is to test this hypothesis in an
independent quantum or stochastic model of gravitational
response. Computing the induced area-change statistics
for a specified matter perturbation would determine
whether agreement with the mean thermodynamic relation
extends to the variance and the full distribution.
Causal diamonds offer a further setting for this
investigation in finite spacetime regions. One could
construct modular-transfer statistics for operations
localized in a diamond and examine their relation to
its boundary-area response, identifying the relevant
modular evolution and the geometric constraints.
This would also permit a more direct comparison with
quantum area fluctuations derived in causal-diamond
models
\cite{CiambelliHeZurek2025,KowalskiGlikmanVarrin2026}.

More broadly, the algebraic formulation provides a
way to connect quantum stochastic thermodynamics
with quantum field theory.
Energy-transfer probabilities, entropy production,
and thermodynamic uncertainty can be studied within
the same framework, including local algebras
that do not admit a density-matrix description.
Developing this connection further may help determine
how the statistical properties of quantum matter
enter horizon thermodynamics beyond mean energy
and entropy balances.
\vspace{0.25cm}

\begin{acknowledgments} 
The author acknowledges financial support from the
S\~ao Paulo Research Foundation (FAPESP), Grant
No.~2026/06852-9, and thanks Nordita for its hospitality
during the development of this work and T. Rick Perche for
helpful discussions.

The author disclose the use of AI tools (OpenAI Codex with GPT-6) to assist in checking the mathematical consistency of the manuscript and reviewing the draft. All ideas, results, and scientific conclusions originate with the author. The author have thoroughly checked the manuscript and take full responsibility for its content.
\end{acknowledgments}

\bibliography{referencias}

@ARTICLE{Crooks1998,
       author = {{Crooks}, Gavin E.},
        title = {Nonequilibrium Measurements of Free Energy Differences for Microscopically Reversible Markovian Systems},
      journal = {Journal of Statistical Physics},
         year = {1998},
        month = {mar},
       volume = {90},
       number = {5-6},
        pages = {1481-1487},
          doi = {10.1023/A:1023208217925}
}

@article{Seifert2012,
doi = {10.1088/0034-4885/75/12/126001},
url = {https://doi.org/10.1088/0034-4885/75/12/126001},
year = {2012},
month = {nov},
volume = {75},
number = {12},
pages = {126001},
author = {Seifert, Udo},
title = {Stochastic thermodynamics, fluctuation theorems and molecular machines},
journal = {Reports on Progress in Physics}
}

@article{Unruh1976,
  author = {Unruh, W. G.},
  title = {Notes on black-hole evaporation},
  journal = {Phys. Rev. D},
  volume = {14},
  pages = {870},
  year = {1976},
  url = {https://doi.org/10.1103/PhysRevD.14.870}
}

@article{Costa2026,
  title = {Work distribution of quantum fields in static curved spacetimes},
  author = {Costa, Rafael L. S. and Basso, Marcos L. W. and Maziero, Jonas and C\'eleri, Lucas C.},
  journal = {Phys. Rev. D},
  volume = {113},
  issue = {2},
  pages = {025010},
  numpages = {10},
  year = {2026},
  month = {Jan},
  publisher = {American Physical Society},
  doi = {10.1103/d26w-cj7z},
  url = {https://link.aps.org/doi/10.1103/d26w-cj7z}
}

@article{Bonfill,
  title = {First law of quantum field thermodynamics},
  author = {Teixid\'o-Bonfill, Adam and Ortega, Alvaro and Mart\'{\i}n-Mart\'{\i}nez, Eduardo},
  journal = {Phys. Rev. A},
  volume = {102},
  issue = {5},
  pages = {052219},
  numpages = {10},
  year = {2020},
  month = {Nov},
  publisher = {American Physical Society},
  doi = {10.1103/PhysRevA.102.052219},
  url = {https://link.aps.org/doi/10.1103/PhysRevA.102.052219}
}

@article{Ortega,
  title = {Work Distributions on Quantum Fields},
  author = {Ortega, Alvaro and McKay, Emma and Alhambra, \'Alvaro M. and Mart\'{\i}n-Mart\'{\i}nez, Eduardo},
  journal = {Phys. Rev. Lett.},
  volume = {122},
  issue = {24},
  pages = {240604},
  numpages = {6},
  year = {2019},
  month = {Jun},
  publisher = {American Physical Society},
  doi = {10.1103/PhysRevLett.122.240604},
  url = {https://link.aps.org/doi/10.1103/PhysRevLett.122.240604}
}

@article{Iso2011,
title = {Non-equilibrium fluctuations of black hole horizons and the generalized second law},
journal = {Phys. Lett. B},
volume = {705},
number = {1},
pages = {152-156},
year = {2011},
issn = {0370-2693},
doi = {10.1016/j.physletb.2011.09.114},
url = {https://www.sciencedirect.com/science/article/pii/S0370269311012147},
author = {Satoshi Iso and Susumu Okazawa and Sen Zhang},
}

@article{Bekenstein74,
  title = {Generalized second law of thermodynamics in black-hole physics},
  author = {Bekenstein, Jacob D.},
  journal = {Phys. Rev. D},
  volume = {9},
  issue = {12},
  pages = {3292--3300},
  numpages = {0},
  year = {1974},
  month = {Jun},
  publisher = {American Physical Society},
  doi = {10.1103/PhysRevD.9.3292},
  url = {https://link.aps.org/doi/10.1103/PhysRevD.9.3292}
}

@article{Hawking1975,
  author = {Stephen W. Hawking},
  title = {Particle Creation by Black Holes},
  journal = {Commun. Math. Phys.},
  volume = {43},
  number = {3},
  pages = {199--220},
  year = {1975},
  doi = {10.1007/BF02345020}
}

@book{Strasberg2022,
  author = {Strasberg, Philipp},
  title = {Quantum Stochastic Thermodynamics: Foundations and Selected Applications},
  publisher = {Oxford University Press},
  year = {2022},
  isbn = {978-0-19-289558-5},
  doi = {10.1093/oso/9780192895585.001.0001},
  url = {https://doi.org/10.1093/oso/9780192895585.001.0001}
}

@article{Hawking71,
  title = {Gravitational radiation from colliding black holes},
  author = {Hawking, S. W.},
  journal = {Phys. Rev. Lett.},
  volume = {26},
  number = {21},
  pages = {1344--1346},
  year = {1971},
  doi = {10.1103/PhysRevLett.26.1344},
  url = {https://doi.org/10.1103/PhysRevLett.26.1344}
}

@article{Bardeen,
  title = {The four laws of black hole mechanics},
  author = {Bardeen, J. M. and Carter, B. and Hawking, S. W.},
  journal = {Commun. Math. Phys.},
  volume = {31},
  number = {2},
  pages = {161--170},
  year = {1973},
  doi = {10.1007/BF01645742},
  url = {https://doi.org/10.1007/BF01645742}
}

@article{Fulling1973,
  title = {Nonuniqueness of canonical field quantization in {Riemannian} space-time},
  author = {Fulling, S. A.},
  journal = {Phys. Rev. D},
  volume = {7},
  number = {10},
  pages = {2850--2862},
  year = {1973},
  doi = {10.1103/PhysRevD.7.2850},
  url = {https://doi.org/10.1103/PhysRevD.7.2850}
}

@article{Bekenstein1973,
  author = {Jacob D. Bekenstein},
  title = {Black Holes and Entropy},
  journal = {Phys. Rev. D},
  volume = {7},
  number = {8},
  pages = {2333--2346},
  year = {1973},
  doi = {10.1103/PhysRevD.7.2333}
}

@article{Crooks1999,
  author = {Gavin E. Crooks},
  title = {Entropy production fluctuation theorem and the nonequilibrium work relation for free energy differences},
  journal = {Phys. Rev. E},
  volume = {60},
  number = {3},
  pages = {2721--2726},
  year = {1999},
  doi = {10.1103/PhysRevE.60.2721}
}

@article{Jarzynski1997,
  author = {Christopher Jarzynski},
  title = {Nonequilibrium equality for free energy differences},
  journal = {Phys. Rev. Lett.},
  volume = {78},
  number = {14},
  pages = {2690--2693},
  year = {1997},
  doi = {10.1103/PhysRevLett.78.2690}
}

@article{Jarzynski11,
   author = {Jarzynski, Christopher},
   title = {Equalities and Inequalities: Irreversibility and the Second Law of Thermodynamics at the Nanoscale}, 
   journal= {Annual Review of Condensed Matter Physics},
   year = {2011},
   volume = {2},
   pages = {329-351},
   doi = {https://doi.org/10.1146/annurev-conmatphys-062910-140506},
   url = {https://www.annualreviews.org/content/journals/10.1146/annurev-conmatphys-062910-140506},
}

@article{Esposito2009,
  author = {Massimiliano Esposito and Upendra Harbola and Shaul Mukamel},
  title = {Nonequilibrium fluctuations, fluctuation theorems, and counting statistics in quantum systems},
  journal = {Rev. Mod. Phys.},
  volume = {81},
  number = {4},
  pages = {1665--1702},
  year = {2009},
  doi = {10.1103/RevModPhys.81.1665},
  url = {https://doi.org/10.1103/RevModPhys.81.1665}
}

@article{Sanders15,
author = {Sanders, Ko},
title = {On the Construction of Hartle–Hawking–Israel States Across a Static Bifurcate Killing Horizon},
journal = {Lett. Math. Phys.},
volume = {105},
pages = {575–640},
year = {2015},
doi = {10.1007/s11005-015-0745-2},
URL = {https://doi.org/10.1007/s11005-015-0745-2},
}

@article{Dappiaggi2009,
    author = {Dappiaggi, Claudio and Moretti, Valter and Pinamonti, Nicola},
    title = {Rigorous construction and {Hadamard} property of the {Unruh} state in {Schwarzschild} spacetime},
    journal = {Adv. Theor. Math. Phys.},
    volume = {15},
    number = {2},
    pages = {355--447},
    year = {2011},
    doi = {10.4310/ATMP.2011.v15.n2.a4}
}

@article{Kay1991,
title = {Theorems on the uniqueness and thermal properties of stationary, nonsingular, quasifree states on spacetimes with a bifurcate killing horizon},
journal = {Phys. Rep.},
volume = {207},
number = {2},
pages = {49-136},
year = {1991},
issn = {0370-1573},
doi = {10.1016/0370-1573(91)90015-E},
url = {https://www.sciencedirect.com/science/article/pii/037015739190015E},
author = {Bernard S. Kay and Robert M. Wald},
}

@book{Haag1996,
  author = {Rudolf Haag},
  title = {Local Quantum Physics: Fields, Particles, Algebras},
  publisher = {Springer},
  year = {1996},
  edition = {2nd}
}

@article{Cai25,
title = {General relativistic fluctuation theorems},
journal = {Phys. Lett. B},
volume = {860},
pages = {139220},
year = {2025},
issn = {0370-2693},
doi = {10.1016/j.physletb.2024.139220},
url = {https://www.sciencedirect.com/science/article/pii/S0370269324007780},
author = {Yifan Cai and Tao Wang and Liu Zhao}
}

@article{Wall2012,
  author = {Aron C. Wall},
  title = {Proof of the generalized second law for rapidly changing fields and arbitrary horizon slices},
  journal = {Phys. Rev. D},
  volume = {85},
  number = {6},
  pages = {104049},
  year = {2012},
  doi = {10.1103/PhysRevD.85.104049}
}

@article{Basso2025,
  title={Quantum detailed fluctuation theorem in curved spacetimes: The observer dependent nature of entropy production},
  author={Basso, M. L. W. and Maziero, J. and C\'{e}leri, L. C.},
  journal={Phys. Rev. Lett.},
  volume={134},
  pages={050406},
  year={2025},
  url = {https://doi.org/10.1103/PhysRevLett.134.050406}
}

@article{Jacobson1995,
  title = {Thermodynamics of Spacetime: The {Einstein} Equation of State},
  author = {Jacobson, Ted},
  journal = {Phys. Rev. Lett.},
  volume = {75},
  pages = {1260--1263},
  year = {1995},
  doi = {10.1103/PhysRevLett.75.1260},
  url = {https://doi.org/10.1103/PhysRevLett.75.1260}
}

@article{Jacobson2016,
  title = {Entanglement Equilibrium and the {Einstein} Equation},
  author = {Jacobson, Ted},
  journal = {Phys. Rev. Lett.},
  volume = {116},
  pages = {201101},
  year = {2016},
  doi = {10.1103/PhysRevLett.116.201101},
  url = {https://doi.org/10.1103/PhysRevLett.116.201101}
}

@article{MassarParentani2000,
  title = {How the change in horizon area drives black hole evaporation},
  author = {Massar, S. and Parentani, R.},
  journal = {Nucl. Phys. B},
  volume = {575},
  pages = {333--356},
  year = {2000},
  url = {https://doi.org/10.1016/S0550-3213(00)00067-5}
}

@article{JacobsonParentani2003,
  title = {Horizon entropy},
  author = {Jacobson, T. and Parentani, R.},
  journal = {Found. Phys.},
  volume = {33},
  pages = {323--348},
  year = {2003},
  url = {https://doi.org/10.1023/A:1023785123428}
}

@article{Eling2006,
  title = {Nonequilibrium Thermodynamics of Spacetime},
  author = {Eling, Christopher and Guedens, Raf and Jacobson, Ted},
  journal = {Phys. Rev. Lett.},
  volume = {96},
  pages = {121301},
  year = {2006},
  doi = {10.1103/PhysRevLett.96.121301},
  url = {https://doi.org/10.1103/PhysRevLett.96.121301}
}

@article{Parikh,
  title = {Einstein's equations from the stretched future light cone},
  author = {Parikh, M. and Svesko, A.},
  journal = {Phys. Rev. D},
  volume = {98},
  issue = {2},
  pages = {026018},
  numpages = {18},
  year = {2018},
  month = {Jul},
  publisher = {American Physical Society},
  doi = {10.1103/PhysRevD.98.026018},
  url = {https://link.aps.org/doi/10.1103/PhysRevD.98.026018}
}

@article{Danielson2022,
  title = {Black holes decohere quantum superpositions},
  author = {Danielson, D. L. and Satishchandran, G. and Wald, R. M.},
  journal = {Int. J. Mod. Phys. D},
  volume = {31},
  issue = {14},
  pages = {2241003},
  numpages = {9},
  year = {2022},
  month = {Oct},
  publisher = {World Scientific},
  doi = {10.1142/S0218271822410036},
  url = {https://doi.org/10.1142/S0218271822410036}
}

@article{Danielson2023,
  title = {Killing horizons decohere quantum superpositions},
  author = {Danielson, D. L. and Satishchandran, G. and Wald, R. M.},
  journal = {Phys. Rev. D},
  volume = {108},
  issue = {2},
  pages = {025007},
  numpages = {16},
  year = {2023},
  month = {Jul},
  publisher = {American Physical Society},
  doi = {10.1103/PhysRevD.108.025007},
  url = {https://link.aps.org/doi/10.1103/PhysRevD.108.025007}
}

@article{GelfandNaimark1943,
  title = {On the imbedding of normed rings into the ring of operators in {Hilbert} space},
  author = {Gelfand, I. and Neumark, M.},
  journal = {Rec. Math. [Mat. Sbornik] N.S.},
  volume = {12(54)},
  number = {2},
  pages = {197--217},
  year = {1943},
  url = {https://www.mathnet.ru/eng/sm6155}
}

@article{Segal1947,
  title = {Irreducible representations of operator algebras},
  author = {Segal, I. E.},
  journal = {Bull. Amer. Math. Soc.},
  volume = {53},
  number = {2},
  pages = {73--88},
  year = {1947},
  doi = {10.1090/S0002-9904-1947-08742-5},
  url = {https://doi.org/10.1090/S0002-9904-1947-08742-5}
}

@article{Borchers2000,
  title = {On revolutionizing quantum field theory with {Tomita}'s modular theory},
  author = {Borchers, H. J.},
  journal = {J. Math. Phys.},
  volume = {41},
  number = {6},
  pages = {3604--3673},
  year = {2000},
  doi = {10.1063/1.533323},
  url = {https://doi.org/10.1063/1.533323}
}

@article{Witten18,
  title = {APS Medal for Exceptional Achievement in Research: Invited article on entanglement properties of quantum field theory},
  author = {Witten, E.},
  journal = {Rev. Mod. Phys.},
  volume = {90},
  issue = {4},
  pages = {045003},
  numpages = {38},
  year = {2018},
  month = {Oct},
  doi = {10.1103/RevModPhys.90.045003},
  url = {https://link.aps.org/doi/10.1103/RevModPhys.90.045003}
}

@book{Takesaki1970,
  title = {{Tomita}'s Theory of Modular {Hilbert} Algebras and its Applications},
  author = {Takesaki, M.},
  series = {Lecture Notes in Mathematics},
  volume = {128},
  publisher = {Springer},
  address = {Berlin, Heidelberg},
  year = {1970},
  doi = {10.1007/BFb0065832},
  url = {https://doi.org/10.1007/BFb0065832}
}

@article{HorowitzGingrich2020,
  title = {Thermodynamic uncertainty relations constrain non-equilibrium fluctuations},
  author = {Horowitz, J. M. and Gingrich, T. R.},
  journal = {Nat. Phys.},
  volume = {16},
  issue = {1},
  pages = {15--20},
  numpages = {6},
  year = {2020},
  month = {Jan},
  publisher = {Springer Science and Business Media LLC},
  doi = {10.1038/s41567-019-0702-6},
  url = {https://doi.org/10.1038/s41567-019-0702-6}
}

@article{Padmanabhan2010,
  title = {Thermodynamical aspects of gravity: new insights},
  author = {Padmanabhan, T.},
  journal = {Rep. Prog. Phys.},
  volume = {73},
  issue = {4},
  pages = {046901},
  numpages = {44},
  year = {2010},
  month = {Apr},
  publisher = {IOP Publishing},
  doi = {10.1088/0034-4885/73/4/046901},
  url = {https://doi.org/10.1088/0034-4885/73/4/046901}
}

@article{Padmanabhan2005,
  title = {Gravity and the thermodynamics of horizons},
  author = {Padmanabhan, T.},
  journal = {Phys. Rep.},
  volume = {406},
  issue = {2},
  pages = {49--125},
  numpages = {77},
  year = {2005},
  month = {Jan},
  publisher = {Elsevier},
  doi = {10.1016/j.physrep.2004.10.003},
  url = {https://doi.org/10.1016/j.physrep.2004.10.003}
}

@article{DeLorenzoPerez2018,
  title = {Light cone thermodynamics},
  author = {De Lorenzo, T. and Perez, A.},
  journal = {Phys. Rev. D},
  volume = {97},
  issue = {4},
  pages = {044052},
  numpages = {19},
  year = {2018},
  month = {Feb},
  publisher = {American Physical Society},
  doi = {10.1103/PhysRevD.97.044052},
  url = {https://link.aps.org/doi/10.1103/PhysRevD.97.044052}
}

@article{BaccettiVisser2014,
  title = {Clausius entropy for arbitrary bifurcate null surfaces},
  author = {Baccetti, V. and Visser, M.},
  journal = {Class. Quantum Grav.},
  volume = {31},
  issue = {3},
  pages = {035009},
  numpages = {20},
  year = {2014},
  month = {Feb},
  publisher = {IOP Publishing},
  doi = {10.1088/0264-9381/31/3/035009},
  url = {https://doi.org/10.1088/0264-9381/31/3/035009}
}

@article{CiambelliHeZurek2025,
  title = {Quantum area fluctuations from gravitational phase space},
  author = {Ciambelli, L. and He, T. and Zurek, K. M.},
  journal = {J. High Energy Phys.},
  volume = {2025},
  number = {08},
  pages = {199},
  year = {2025},
  url = {https://doi.org/10.1007/JHEP08(2025)199}
}

@article{KowalskiGlikmanVarrin2026,
  title = {Quantum geometry from area fluctuations},
  author = {Kowalski-Glikman, J. and Varrin, L.},
  journal = {J. High Energy Phys.},
  volume = {2026},
  number = {08},
  pages = {007},
  year = {2026},
  url = {https://doi.org/10.1007/JHEP08(2026)007}
}

@article{ChircoLiberati2010,
  title = {Non-equilibrium thermodynamics of spacetime: The role of gravitational dissipation},
  author = {Chirco, Goffredo and Liberati, Stefano},
  journal = {Phys. Rev. D},
  volume = {81},
  pages = {024016},
  year = {2010},
  doi = {10.1103/PhysRevD.81.024016},
  url = {https://doi.org/10.1103/PhysRevD.81.024016}
}

@article{BianchiSatz2013,
  title = {Mechanical laws of the {Rindler} horizon},
  author = {Bianchi, Eugenio and Satz, Alejandro},
  journal = {Phys. Rev. D},
  volume = {87},
  pages = {124031},
  year = {2013},
  doi = {10.1103/PhysRevD.87.124031},
  url = {https://doi.org/10.1103/PhysRevD.87.124031}
}

@article{DorauMuch2026,
  title = {From Quantum Relative Entropy to the Semiclassical {Einstein} Equations},
  author = {Dorau, Philipp and Much, Albert},
  journal = {Phys. Rev. Lett.},
  volume = {136},
  pages = {091602},
  year = {2026},
  doi = {10.1103/lmq8-nsty},
  url = {https://doi.org/10.1103/lmq8-nsty}
}

@article{IsoOkazawa2011,
  title = {Stochastic equations in black hole backgrounds and non-equilibrium fluctuation theorems},
  author = {Iso, Satoshi and Okazawa, Susumu},
  journal = {Nucl. Phys. B},
  volume = {851},
  pages = {380--419},
  year = {2011},
  doi = {10.1016/j.nuclphysb.2011.05.021},
  url = {https://doi.org/10.1016/j.nuclphysb.2011.05.021}
}

@article{CiraficiNESS2024,
  title = {On the nonequilibrium dynamics of gravitational algebras},
  author = {Cirafici, Michele},
  journal = {Class. Quantum Grav.},
  volume = {41},
  number = {23},
  pages = {235006},
  year = {2024},
  doi = {10.1088/1361-6382/ad85bf},
  url = {https://doi.org/10.1088/1361-6382/ad85bf}
}

@article{CiraficiFT2024,
  title = {Fluctuation theorems, quantum channels and gravitational algebras},
  author = {Cirafici, Michele},
  journal = {J. High Energy Phys.},
  volume = {2024},
  number = {11},
  pages = {089},
  year = {2024},
  doi = {10.1007/JHEP11(2024)089},
  url = {https://doi.org/10.1007/JHEP11(2024)089}
}

@article{Takeda2026,
  title = {Work distribution and fluctuation theorem in {AdS/CFT}},
  author = {Takeda, Daichi},
  journal = {Phys. Rev. D},
  volume = {113},
  pages = {106029},
  year = {2026},
  doi = {10.1103/6hw6-t8l7},
  url = {https://doi.org/10.1103/6hw6-t8l7}
}

@misc{BassoSaa2026,
  title = {Heat distribution of quantum fields interacting with {Unruh--DeWitt} detectors},
  author = {Basso, Marcos L. W. and Saa, Alberto},
  year = {2026},
  eprint = {2609.06185},
  archivePrefix = {arXiv},
  primaryClass = {gr-qc}
}

@incollection{JOPP2011,
  title = {Entropic fluctuations in quantum statistical mechanics---an introduction},
  author = {Jak{\v s}i{\'c}, Vojkan and Ogata, Yoshiko and Pautrat, Yan and Pillet, Claude-Alain},
  booktitle = {Quantum Theory from Small to Large Scales},
  editor = {Fr{\"o}hlich, J{\"u}rg and Salmhofer, Manfred and Mastropietro, Vieri and De Roeck, Wojciech and Cugliandolo, Leticia F.},
  series = {Lecture Notes of the Les Houches Summer School},
  volume = {95},
  pages = {213--410},
  publisher = {Oxford University Press},
  address = {Oxford},
  year = {2012},
  doi = {10.1093/acprof:oso/9780199652495.003.0004},
  url = {https://doi.org/10.1093/acprof:oso/9780199652495.003.0004}
}

@article{Guedens2012,
doi = {10.1088/0264-9381/29/14/145002},
url = {https://doi.org/10.1088/0264-9381/29/14/145002},
year = {2012},
month = {jun},
volume = {29},
number = {14},
pages = {145002},
author = {Guedens, R.},
title = {Locally inertial null normal coordinates},
journal = {Class. Quantum Grav.}
}

@article{Guedens2012a,
  title = {Horizon entropy and higher curvature equations of state},
  author = {Guedens, R. and Jacobson, T. and Sarkar, S.},
  journal = {Phys. Rev. D},
  volume = {85},
  issue = {6},
  pages = {064017},
  numpages = {16},
  year = {2012},
  doi = {10.1103/PhysRevD.85.064017},
  url = {https://link.aps.org/doi/10.1103/PhysRevD.85.064017}
}

@article{JOPS2012,
  title = {Quantum Hypothesis Testing and Non-Equilibrium Statistical Mechanics},
  author = {Jak{\v s}i{\'c}, Vojkan and Ogata, Yoshiko and Pillet, Claude-Alain and Seiringer, Robert},
  journal = {Rev. Math. Phys.},
  volume = {24},
  pages = {1230002},
  year = {2012},
  doi = {10.1142/S0129055X12300026},
  url = {https://doi.org/10.1142/S0129055X12300026}
}

@article{Araki1976,
  title = {Relative entropy of states of von {Neumann} algebras},
  author = {Araki, Huzihiro},
  journal = {Publ. Res. Inst. Math. Sci.},
  volume = {11},
  pages = {809--833},
  year = {1976},
  doi = {10.2977/prims/1195191148},
  url = {https://doi.org/10.2977/prims/1195191148}
}

@article{Longo2019,
  title = {Entropy of coherent excitations},
  author = {Longo, Roberto},
  journal = {Lett. Math. Phys.},
  volume = {109},
  pages = {2587--2600},
  year = {2019},
  doi = {10.1007/s11005-019-01196-6},
  url = {https://doi.org/10.1007/s11005-019-01196-6}
}

@article{CasiniGrilloPontello2019,
  title = {Relative entropy for coherent states from {Araki} formula},
  author = {Casini, Horacio and Grillo, Sergio and Pontello, Diego},
  journal = {Phys. Rev. D},
  volume = {99},
  pages = {125020},
  year = {2019},
  doi = {10.1103/PhysRevD.99.125020},
  url = {https://doi.org/10.1103/PhysRevD.99.125020}
}

@misc{GallockHasegawa2026,
  title = {Trade-off Relation for Black Hole Entropy Fluctuations},
  author = {Gallock-Yoshimura, Kensuke and Hasegawa, Yoshihiko},
  year = {2026},
  eprint = {2605.26214},
  archivePrefix = {arXiv},
  primaryClass = {gr-qc}
}

@article{SummersVerch1996,
  author  = {Summers, Stephen J. and Verch, Rainer},
  title   = {Modular inclusion, the {Hawking} temperature, and quantum field theory in curved spacetime},
  journal = {Lett. Math. Phys.},
  volume  = {37},
  number  = {2},
  pages   = {145--158},
  year    = {1996},
  doi     = {10.1007/BF00416017}
}

@article{BaratoSeifert2015,
  author  = {Barato, Andre C. and Seifert, Udo},
  title   = {Thermodynamic uncertainty relation for biomolecular processes},
  journal = {Phys. Rev. Lett.},
  volume  = {114},
  number  = {15},
  pages   = {158101},
  year    = {2015},
  doi     = {10.1103/PhysRevLett.114.158101}
}

@article{BisognanoWichmann1975,
  author  = {Bisognano, Joseph J. and Wichmann, Eyvind H.},
  title   = {On the duality condition for a {Hermitian} scalar field},
  journal = {J. Math. Phys.},
  volume  = {16},
  number  = {4},
  pages   = {985--1007},
  year    = {1975},
  doi     = {10.1063/1.522605}
}

@article{KPV2021,
  title = {Temperature and entropy--area relation of quantum matter near spherically symmetric outer trapping horizons},
  author = {Kurpicz, Fiona and Pinamonti, Nicola and Verch, Rainer},
  journal = {Lett. Math. Phys.},
  volume = {111},
  pages = {110},
  year = {2021},
  doi = {10.1007/s11005-021-01445-7},
  url = {https://doi.org/10.1007/s11005-021-01445-7}
}
\end{document}